\documentclass[journal]{article}                                  

\usepackage{arxiv}
\usepackage{hyperref}
\usepackage{graphicx}          
\usepackage{amsmath} 
\usepackage{amssymb}  

\usepackage{amsthm}
\usepackage{color,soul}
\usepackage{flushend}
\usepackage{cite}
\usepackage{mfirstuc}
\usepackage{mathtools}
\usepackage{xcolor}
\usepackage{array,multirow}
\usepackage{algorithm} 
\usepackage{algorithmic}  
\usepackage[linesnumbered,algo2e,ruled,vlined,norelsize]{algorithm2e} 

\allowdisplaybreaks
\usepackage{tikz}
\usepackage{textcomp}
\usepackage{hyperref}
\usepackage{lipsum}

\newcommand\copyrighttext{%
  \footnotesize \textcopyright  This work is under review.
  Permission from the authors must be obtained for all other uses, in any current or future 
  media, including reprinting/republishing this material for advertising or promotional 
  purposes, creating new collective works, for resale or redistribution to servers or 
  lists, or reuse of any copyrighted component of this work in other works. 
  DOI: \href{<http://tex.stackexchange.com>}{To be generated.}}
\newcommand\copyrightnotice{%
\begin{tikzpicture}[remember picture,overlay]
\node[anchor=south,yshift=10pt] at (current page.south) {\fbox{\parbox{\dimexpr\textwidth-\fboxsep-\fboxrule\relax}{\copyrighttext}}};
\end{tikzpicture}%
}

\usepackage{authblk}
\title{\LARGE \bf
Secure Estimation of Battery Voltage Under Sensor Attacks:\\ A Self-Learning Koopman Approach
}

\author[1]{Sanchita Ghosh}
\author[1]{Tanushree Roy}
\affil[1]{Department of  Mechanical Engineering, Texas Tech University, Lubbock, TX 79409, US. Emails:~{\tt\small sancghos@ttu.edu, tanushree.roy@ttu.edu}.}

\begin{document}

\copyrightnotice

\maketitle
\thispagestyle{empty}
\pagestyle{empty}


\begin{abstract}
Cloud-based battery management system (BMS) requires accurate terminal voltage measurement data to ensure optimal and safe charging of Lithium-ion batteries. Unfortunately, an adversary can corrupt the battery terminal voltage data as it passes from the local-BMS to the cloud-BMS through the communication network, with the objective of under- or over-charging the battery. To ensure accurate terminal voltage data under such malicious sensor attacks, this paper investigates a Koopman-based secure terminal voltage estimation scheme using a two-stage error-compensated self-learning feedback.  During the first stage of error correction, the potential Koopman prediction error is estimated to compensate for the error accumulation due to the linear approximation of Koopman operator. The second stage of error compensation aims to recover the error amassing from the higher-order dynamics of the Lithium-ion batteries missed by the self-learning strategy. Specifically, we have proposed two different methods for this second stage error compensation. First, an interpretable empirical correction strategy has been obtained using the open circuit voltage to  state-of-charge mapping for the battery. Second, a Gaussian process regression-based data-driven method has been explored.  Finally, we demonstrate the efficacy of the proposed secure estimator using both empirical and data-driven corrections.
\end{abstract}

\section{Introduction}
Renewable energy systems are essential for tackling the growing climate change challenges caused by conventional fuels \cite{amir2023energy}. Lithium-ion batteries have become the preferred choice for energy storage and a reliable power source for various applications, including electric vehicles, to accommodate this transition toward cleaner energy sources \cite{stan2014lithium}. Similarly, BMS play a critical role in enhancing the efficiency, safety, and lifespan of these batteries by real-time monitoring and regulation of key battery states such as the state of charge, temperature, or terminal voltage. Large-scale deployment and co-ordination of these batteries also require the BMS to operate in a cyber-physical environment i.e. the BMS is implemented in the cloud to leverage high computational resources \cite{kim2020overview}. Unfortunately, such cyber-physical BMS operation have inherent security risks that can result in financial and energy losses or even unsafe operations. This necessitates the research on cyber-security analysis for BMS to ensure reliable and safe battery operations against cyberattacks \cite{ghosh2023security,naseri2023cyber}. %

To address these challenges, \cite{murlidharan2025battery} provided a comprehensive threat assessment to identify the potential impact of cyberattacks on the functionality, safety, and performance of BMS. Furthermore,  authors proposed  a extended Kalman filter-based algorithm to detect false-data-injection sensor attacks in battery stacks in  \cite{obrien2023detection}. In \cite{dey2020cybersecurity}, authors investigated model-based strategies to detect concurrent actuation and sensor cyberattacks for electric vehicle (EV) battery systems. Similarly, machine learning models such  deep neural network \cite{basnet2020deep} and supervised classifier \cite{elkashlan2023machine} have been implemented to detect attacks on EV charging. Moreover, the authors adopted a Koopman operator-based data-driven approach to detect both actuation and sensor attacks and thereafter, isolate the source of the attack in \cite{ghosh5028845detection}.  On the other hand, in \cite{khalid2021investigation} utilized the Principal Component Analysis based unsupervised k-means approach to detect the presence of cell-voltage buffer manipulation cyberattacks for BMS.

Beyond detection of these attacks, it is crucial to develop countermeasure strategies to mitigate the impact of cyberattacks upon detection \cite{roy2024input}. While actuation attacks can directly impact the system causing severe outcomes, their impact can be detected more readily and mitigated through safety constraints or actuation removal. In contrast, sensor attacks are designed to deceive administrators about the true state of the system and in general remain harder to detect and mitigate without sensor redundancy. Thus, failure to mitigate sensor attacks can lead to incorrect control actions and unpredictable failures. Accordingly, authors proposed a non-linear estimator based on equivalent-circuit model (ECM) for Li-ion batteries to ensure secure state-of-charge (SOC) estimation under sensor cyberattacks in \cite{tian2022security}. Furthermore, authors utilized the extended Kalman filter method to identify the parameters for the ECM. In \cite{wang2022secure}, authors investigated a time-varying recursive Kalman filter to ensure secure SOC estimation upon detection for low-intensity deception attacks.  Moreover, \cite{xiao2025resource} proposed a sensor attack detection and compensation scheme for EV battery systems utilizing the discrepancy in set-membership of the state estimation and prediction to estimate the state of charge of the battery under sensor attacks. 

However, these studies reveal several key research gaps. First, there is a need for secure estimation of battery voltage under corrupt voltage sensing. Second,  we seek a secure estimation method that minimizes the dependence on model knowledge while avoiding the need for large data. Third, the estimation method must eliminate all assumptions regarding the type of cyberattack. Our contribution addresses these gaps in the following way: 
\begin{enumerate}
    \item We proposed a self-learning feedback mechanism for Koopman operator based secure terminal voltage estimation for Lithium-ion batteries under sensor cyberattak. The self-learning mechanism is assisted by two-staged error compensations that eliminate errors accrued from Koopman approximation and the lack of dynamics data from the battery. The proposed secure estimator adopts a small data online learning approach and only requires limited model information. It is also adaptable to changes in battery dynamics, e.\,g., due to aging, temperature, or operation conditions.
    \item For the error correction, we propose an interpretable battery characteristic guided empirical correction for self-learning Koopman operator. 
    \item Alternatively, we also propose a GPR-based error correction for self-learning Koopman operator that requires minimal offline data for training and eliminates the need for battery characteristic information.
\end{enumerate}

The rest of the paper is organized as follows. Section \ref{probForm} presents the problem framework of this paper. Next, we introduce our proposed self-learning Koopman-based secure estimation algorithm. Our simulation results are presented in Section \ref{sim}. Finally, Section \ref{conclu} concludes our paper.


\section{Problem Framework: Sensor Attacks During Electric Vehicle Charging} \label{probForm}
This paper addresses the problem of secure voltage estimation during sensor cyberattacks on a Lithium-ion battery while charging. Secure estimation is crucial in this scenario since inaccurate voltage measurement can lead to poor battery management. Sensor attacks on batteries are a critical risk especially for  EV charging, where sensor attacks can occur through the electric vehicle supply equipment (EVSE) at smart charging stations  \cite{johnson2022cybersecurity}. Here, a cloud-based BMS/controller monitors the terminal voltage measurement of the EV battery from the local-BMS to generate appropriate charging current actuation input to the local-BMS such that optimum battery charging and grid stability are ensured. Both the measurement from the local-BMS to the cloud-BMS and the actuation command from the cloud-BMS to the local-BMS are sent through communication channels that are susceptible to cyberattacks. The corruption of the former is known as the sensor attack, while the latter is known as the actuation attack \cite{johnson2022cybersecurity}. 
In this paper, we consider that an adversary injects a sensor cyberattack to corrupt the voltage measurement and the battery charging dynamics in the presence of sensor attack $\delta$ is defined as follows:
\begin{align} \label{vt_dynamics}
      x(k+1) = f(x(k),I_c(k)); \quad V(k) = g(x(k)) + \delta.
\end{align}
\noindent
Here, $x \in \mathbb{R}^3$ represents the battery states of a second order equivalent-circuit battery model, i.\,e., SOC and two voltages across the two resistance-capacitor couple \cite{samad2014parameterization}. $I \in \mathbb{R}$ is the operating current, and the non-linear function $f: \mathbb{R}^d \times \mathbb{R} \rightarrow \mathbb{R}^d$ captures the battery internal dynamics. $V \in \mathbb{R}$ is the terminal voltage measurement and $g: \mathbb{R}^d  \rightarrow \mathbb{R}$ defines the non-linear measurement function. 

\begin{figure}[h!]
    \centering
    \includegraphics[width=0.8\linewidth]{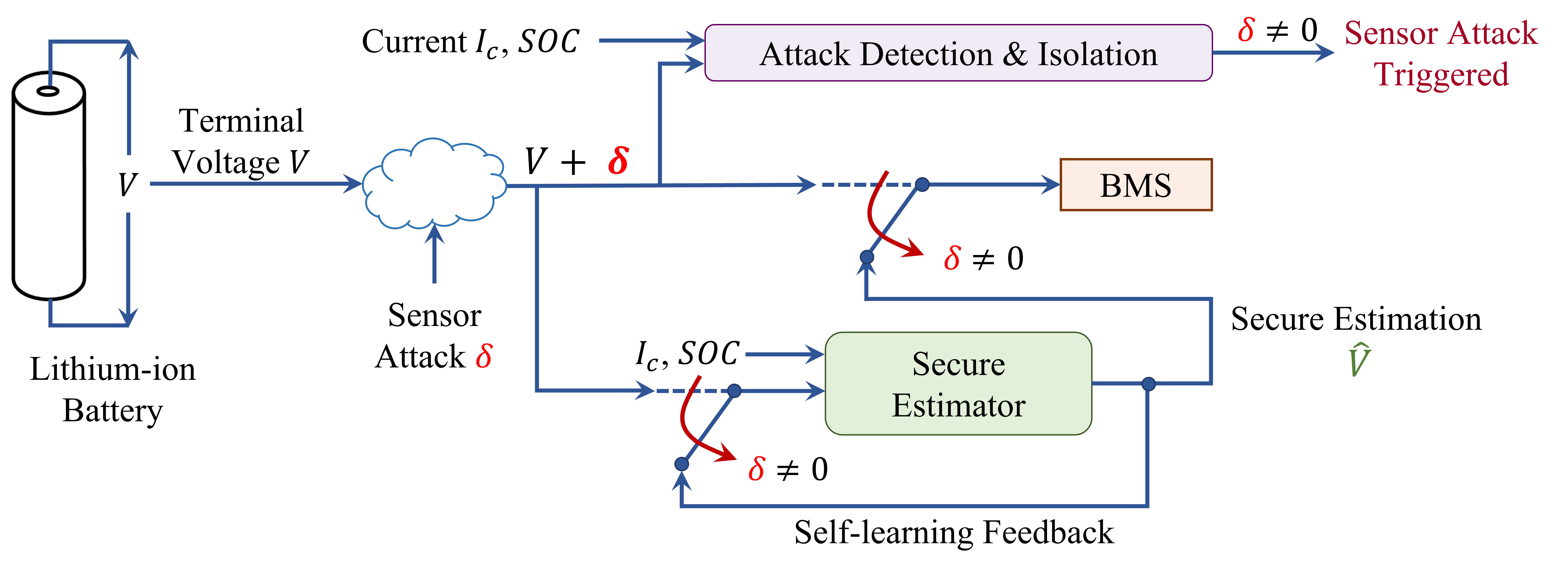}
    \caption{Block diagram shows  the secure terminal voltage estimation generation under sensor attack.}
    \label{fig:overallB}
\end{figure}

In this framework, we propose a event-triggered secure estimator to obtain secure terminal voltage estimation for the Lithium-ion battery subjected to malicious sensor attack. In particular, we utilize the attack detection-isolation algorithm (proposed in our previous paper \cite{ghosh2024koopman})  to reliably activate the secure estimator under sensor attack. Furthermore, we adopt a self-learning Koopman  approach for the proposed secure estimator.  Consequently, the secure estimator runs with error-compensated self-learning feedback when $\delta \neq 0$. Finally, the generated secure terminal voltage estimation is sent to the BMS.  Fig.~\ref{fig:overallB} captures the overview of our problem framework. 

 \section{Secure Terminal Voltage Estimation Algorithm} \label{algo}
In this section, we propose our error compensation integrated self-learning  KO-based secure estimation algorithm. This estimation method has two parts: (A) Koopman-based voltage prediction using self-learning feedback and (B) two-stage prediction error corrections. The second stage of error correction can be implemented using an interpretable empirical strategy based on open circuit voltage (OCV) to SOC mapping or using the Gaussian process regression method.

\subsection{Koopman-based voltage prediction using self-learning feedback}
To learn the nonlinear and complex battery dynamics online, we adopt the Koopman operator-based data-driven approach that maps the finite dimensional nonlinear model  \eqref{vt_dynamics} to an infinite-dimensional linear model in the space of observables. However, using computationally efficient algorithms, such as Hankel Dynamic Mode Decomposition (HDMD) and delay embedding, we can obtain a finite-dimensional approximation of the Koopman operator using limited data \cite{brunton2019notes}.  In this work, we consider that the terminal voltage measurement $V_t$, current input $I_c$, the battery capacity $Q$, and open circuit voltage (OCV) vs the state of charge (SOC) map are known to us.  We utilized the input current data $I_c$ and battery capacity $Q$ to calculate SOC with Coulomb counting, i.\,e., $SOC ({k+1}) = SOC(k) + (\Delta t\,   I_c)/Q$ . The current data $I_c$ and the calculated SOC are used as inputs to our Koopman linear model such that $u(k) = \begin{bmatrix}
    I_c & SOC(k)
\end{bmatrix}^T$. 

To enable small data learning and computational efficiency, the Koopman operator is learned online over a sliding window of voltage data. During each sliding window, we utilize the available data to learn the Koopman linear model and this learned model is utilized to generate terminal voltage prediction $V_p$ over a receding horizon.  Hence, we define a sliding window sequence of $S$ observations, where the learning window length $  \Tilde{S} < S$ to obtain the two data sub-sequences: 
\begin{align}
    & \text{Learning window:} \, \, \quad \, \mathfrak{L} \in \{ k+1, \cdots, k+\Tilde{S}\}, \label{lw}\\
    & \text{Prediction window:} \quad \mathfrak{P} \in \{ k+ \Tilde{S}+1, \cdots, k+ S\}. \label{pw}
\end{align}
 We note here that the learning window $\mathfrak{L}$ is larger than the prediction window $\mathfrak{P}$ and the sliding window is moved ahead with $\overline{S} =  S- \Tilde{S}-1$ amount after every prediction cycle. Using the battery data from the learning window, we obtain the approximate Koopman linear model as follows:
\begin{align}\label{z_app}
   & z(k+1) = A_{\mathfrak{L}} z(k) + B_{\mathfrak{L}} u(k);  \\ & {V_p}(k+1) \approx C_{\mathfrak{L}} z(k+1),
\end{align}
where  $A_{\mathfrak{L}}$,  $B_{\mathfrak{L}}$, and $C_{\mathfrak{L}}$ provide us the Koopman linear approximation for the battery system \eqref{vt_dynamics} from the limited available measurement and input data and these matrices are re-learned at each sliding window. ${V_p}(k)$ is the predicted terminal voltage at the $k$-th time instant. 
However, when a sensor attack triggers the secure estimation algorithm, corrupted terminal voltage measurements are excluded from updating the Koopman operator. Instead, the secure estimator starts to feedback the Koopman estimation $V_p$ to itself, aka initiating a self-learning feedback. 
Consequently,  after some time, the secure estimator utilizes a data stack consisting of only voltage predictions. Thus, under nominal EV charging ($\delta = 0$), the Koopman voltage predictor runs with measurement feedback from the data stack $\zeta_V$ in \eqref{ko_est_v0}. In the presence of a sensor attack ($\delta \neq 0)$, the Koopman voltage predictor uses self-learning feedback using the data stack $\zeta_{{V_p}}$ in \eqref{ko_est_v1}. 
 \begin{align}
     &\zeta_{V} = \begin{bmatrix}
         V(k)  &  \cdots  & {V} ({k + \Tilde{S} })
     \end{bmatrix}; \label{ko_est_v0} \\
      & \zeta_{{V_p}} = \begin{bmatrix}
         {V}_p(k)  &  \cdots  & {{V_p}({k + \Tilde{S})} } 
     \end{bmatrix}. \label{ko_est_v1}
 \end{align}
 In contrast, for both nominal and secure estimation operation,  the Koopman voltage predictor utilizes the input data stack $\zeta_u =  \begin{bmatrix}
         u(k)  &  \cdots  & {u} ({k + S })
     \end{bmatrix}$.
     
Now, if the secure estimator receives only $V_p$ for self-learning feedback, it will accumulate prediction errors from Koopman approximation and higher-order changes in the system dynamics. 
Next, we propose two-stage error corrections for the $V_p$ feedback to address the inaccuracies due to Koopman approximation (stage I) and higher-order battery dynamics (stage II).

\subsection{Prediction \& correction of Koopman estimation error}

 \noindent
 \subsubsection{Stage I: Koopman approximation error correction} 
 Firstly, we consider compensating for the potential prediction error due to Koopman linear approximation.
 Thus, we estimate the potential prediction error $E_1$ such that $E_1 \approx V_{nom} - V_p$, where $V_{nom}$ denotes the nominal voltage measurement. The latter is only available during nominal operation and thus after the sensor attack, this Koopman approximation error is similarly self-learned.  Next,   the predicted voltage $V_p$ after stage I correction $E_1$ is denoted as $\overline{V}_p = V_p  + E_1$. 
 
 The self-learned approximation error compensation while largely improves the self-learned terminal voltage estimation, it still fails to capture the higher-order batttery dynamics. Thus, we add stage II compensation using both empirical strategy and data-driven strategy.

\noindent
 \subsubsection{Stage II: Higher-order dynamics Correction} We have proposed two different methods for this stage. The first method is proposed based on our empirical findings and the second one adopts a data-driven approach.

With this second stage correction, we update the self-learned secure estimator with a comprehensive error compensated estimation $\widehat{V}$ instead of voltage prediction $V_p$ in \eqref{ko_est_v1}, where $\widehat{V} =  V_p + E_2$ and $E_2$ is the comprehensive error compensation using the stage~I error $E_1$.

\begin{figure}[t!]
    \centering
    \includegraphics[width=0.95\linewidth]{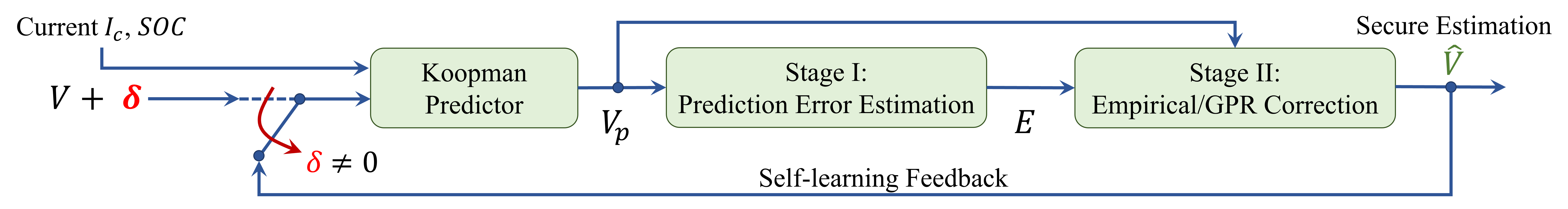}
    \caption{Block diagram showing the components of the secure estimator along with the sensor attack triggered self-learning feedback mechanism.}
    \label{fig:pic2}
\end{figure}

\begin{algorithm2e} 
\caption{{Generate \texttt{Secure Terminal Voltage Estimation} }}\label{alg:desicion}
\KwIn{{Time instant $k$, terminal voltage  measurements $V$, charging current input $I$, calculated $SOC$, choice of  corrector $\mu$, empirical functions $l^j_{i}, m^{j|j-1}_{i}$, pre-trained GPR models $\mathcal{G}_j$.}}
\KwOut{ \texttt{Secure voltage estimation $\widehat{V}$}.}
\SetKwFunction{KLM}{Koopman Predictor}
\SetKwFunction{SE}{Secure Estimator}

\SetKwFunction{EC}{Empirical Corrector }
\SetKwFunction{GA}{GPR Corrector}

    \For{$k\geqslant0$}{
   
    \eIf{$\delta \neq 0 $}{
    $[ \widehat{V}] \leftarrow$  \texttt{Secure Estimator $\left( \zeta_{\widehat{V}}, \zeta_u  \right) $} \\
     \KwRet {\texttt{Secure Estimation $\widehat{V}$}   \;}
    }
    { $[ V_p] \leftarrow$  \texttt{Koopman Predictor $\left( \zeta_V, \zeta_u  \right) $} \,\,\,
    \KwRet {\texttt{Voltage Prediction $V_p$ } \;} } 
    }
    \SetKwProg{Fn}{function}{:}{\KwRet}
    \Fn{\KLM{$\zeta_V, \zeta_u  $}} {Evaluate $V_p$ \eqref{z_app}; \,\,
        \KwRet {$V_p$ \;}
        }
    \SetKwProg{Fn}{function}{:}{\KwRet}
    \Fn{\SE{$\zeta_{\widehat{V}}, \zeta_u  $}}{
        
        \eIf{$\mu == $ "Empirical"}{
        $[ V_p]  \leftarrow$  \texttt{Koopman Predictor $\left( \zeta_{\widehat{V}}, \zeta_u  \right) $};\\
        Find prediction error estimation $E_1$; \\
         $[ \widehat{V}] \leftarrow$  \texttt{Empirical Corrector $\left( V_p , E_1 , l_{i,j} \right) $}; \\
        }{ $[ V_p]  \leftarrow$  \texttt{Koopman Predictor $\left( \zeta_{\overline{V}_p}, \zeta_u  \right) $};\\
        Find prediction error estimation $E_1$; \\
        $[ \widehat{V}]  \leftarrow$  \texttt{GPR Corrector $\left( V_p, E_1, \zeta_u , \mathcal{G}_j \right) $};}
      
        \KwRet {\texttt{Secure Estimation $\widehat{V}$}   \;}
        }
        \SetKwProg{Fn}{function}{:}{\KwRet}
    \Fn{\EC{$ V_p , E_1 , l_{i,j}  $}} {Calculate $E_2$ using \eqref{total_compen}, \eqref{eee}; \\
        Evaluate $\widehat{V} = V_p + E_2$\; \KwRet {$\widehat{V}$ \;}
        }
        \SetKwProg{Fn}{function}{:}{\KwRet}
    \Fn{\GA{$ V_p, E_1, \zeta_u, \mathcal{G}_j  $}} {Predict $\widehat{V}$ with $\mathcal{G}_j$\; \KwRet {$\widehat{V}$ \;}
        }
\end{algorithm2e}
\noindent
{\textbf{Empirical correction with OCV-SOC mapping}}

\noindent
 For this first method, we propose an empirically derived strategy for the error compensation to improve the accuracy of our voltage prediction during self-learning Koopman-based voltage estimation. Our goal in empirically correcting the error compensation $E_2$ is to capture the higher-order dynamics in true terminal voltage $\Delta V_{nom}$ based on the charging current. From the physics-based battery models such as equivalent circuit models, we know that the terminal voltage has a strong relation with the battery OCV through the current or SOC. 
 Therefore,  we consider that higher-order dynamics in terminal voltage can be captured by tracking the higher-order characteristic changes in OCV with corresponding changes in SOC. However, to reduce dependencies on system knowledge, we consider only tracking the changes in OCV using the OCV-SOC map. Most importantly, the OCV-SOC map is comparatively more easily accessible and does not vary with factors such as aging or operating conditions except for the cell chemistry. 

  Our empirical analysis shows that the secure estimator accuracy without a stage II compensation varies through the different regions of SOC. This finding is also congruent with the changes OCV-SOC map. Thus, we propose the following algebraic formula to obtain the empirical correction:
 \begin{align} \label{total_compen}
    E_2 =  l^j_{1} (SOC) E_1 & + l^j_{2} (SOC) \Delta OCV \nonumber  \\
    &+ l^j_{3} (SOC) \Delta^2 OCV.
 \end{align}
Here, $l^j_{1}$, $l^j_{2}$, and $l^j_{3}$ are the SOC-dependent functions such that they exhibit a varied mapping to SOC in $j$-th region of SOC, i.e., $SOC_j$.
  Here $\Delta OCV = OCV(k+1) - OCV(k) $ in \eqref{total_compen} captures the changes in OCV with charging and   $\Delta^2 OCV = \Delta OCV(k+1) - \Delta OCV(k) $ captures the changes in OCV slope. The specific functional forms for these functions for different SOC regions are shown in Table~\ref{tab:ls}. 

Next, the self-learning based prediction error $E_1$ in \eqref{total_compen} also remains relatively constant over a specific SOC region, as it similarly lacks the higher-order dynamics of the battery model. Thus, the prediction error $E^{j-1}_1$ is updated to $E^{j}_1$ when the battery enters a new SOC region $j$ from $j-1$. This empirical compensation of error is denoted by:

\begin{align}\label{eee}
    {E}^{j}_1 = & m^{j|j-1}_{1} (SOC) E^{j-1}_1 + m^{j|j-1}_{2} (SOC_j) \Delta OCV \nonumber \\
     &+ m^{j|j-1}_{3} (SOC_j) \Delta^2 OCV.
\end{align}
The specific functional forms for these functions $m^{j|j-1}_i, \forall i=\{1,2,3\}$ for different SOC region changes are shown in Table~\ref{tab:ms}.

\vspace{2mm}
\noindent
\textbf{Data-driven correction with GPR}

\noindent
For this second method, we adopt a data-driven approach utilizing Gaussian process regression (GPR) to directly approximate the comprehensive error compensation $E_2$. GPR is a non-parametric supervised regression technique that works based on the Gaussian process concept. Gaussian processes are a collection of random variables such that every finite collection of those random variables has a multivariate normal distribution. Thus, GPR attempts to find the underlying structure of the data points and provides predictions as probability distributions. Moreover, GPR is particularly effectively at dealing with problems involving continuous data when the relationship between input and output variables is unknown or complex.   Thus, we choose GPR to approximate the comprehensive error $E_2$ to obtain the secure error-compensated terminal voltage estimation $\widehat{V}$.

To generate training data for the GPR, we first simulate the Koopman estimator with only prediction error corrected self-learning feedback, i.\,e., $\zeta_{\overline{V}_p} = \begin{bmatrix}
         {\overline{V}_p}(k)  &  \cdots  & {{\overline{V}_p} ({k + \Tilde{S}}) } 
     \end{bmatrix}$, here  $\overline{V}_p = V_p  + E_1$. Next, we store the generated terminal voltage prediction $V_p$ with the self-learning feedback $\zeta_{\overline{V}_p}$. 
We consider the charging current $I$, calculated SOC, the predicted terminal voltage $\overline{V}_p$ under $\zeta_{\overline{V}_p}$-based self-learning and the corresponding prediction error estimation $E_1$ data as input to our GPR model. 
In addition, we consider the corresponding nominal terminal voltage $V_{nom}$ to obtain $\overline{V}_p-V_{nom}$ as the output of GPR. We trained separate GPR models $\mathcal{G}_j$ in the $j$-th SOC region, as described in Table~\ref{tab:ls}, to obtain the most accurate voltage estimates. Finally, we utilize these pre-trained GPR models comprehensive error predictions to generate secure voltage estimation $\widehat{V}$.

\section{Simulation Case Study} \label{sim}

In this study, we consider a nonlinear second-order ECM for a prismatic cell Nickel-Manganese-Cobalt (NMC) Lithium-ion battery for  the nominal terminal voltage measurement data generation  \cite{samad2014parameterization}. The $1^{st}$ plot in Fig~\ref{fig:soc_ocv} presents the OCV-SOC curve for this battery. The $2^{nd}$ and the $3^{rd}$ plot shows  $\Delta OCV$ (middle), and $\Delta^2 OCV$ graphs respectively. In addition, the dotted line in each plot marks the different SOC region considered for our empirical correction. Furthermore, the {SOC-dependent functions required for empirical correction of the voltage prediction and of the error in voltage prediction are respectively listed in Table~\ref{tab:ls} and Table~\ref{tab:ms}.  Next, we present two case study for this battery to demonstrate the efficacy of the proposed secure estimator. For both case studies, we consider that the battery is charged with  a $5 A$ current under the CCCV policy and with a initial SOC of $35\%$. 

\begin{figure}
      \centering
      \includegraphics[width=0.65\linewidth]{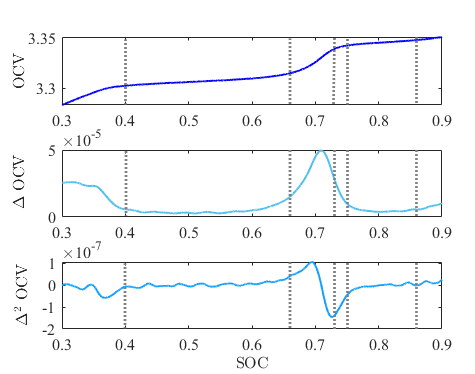}
      \caption{For the prismatic cell NMC Lithium-ion battery, the  figure exhibits OCV-SOC  (top), $\Delta OCV$ (middle), and $\Delta^2 OCV$ graphs. }
      \label{fig:soc_ocv}
  \end{figure}

\renewcommand{\arraystretch}{1.2}
  \begin{table}[h]
      \centering
      \begin{tabular}{|l|c|c|c|}
      \hline
          SOC Regions $j$ & $l^j_{1}$ & $l^j_2$ & $l^j_3$ \\
          \hline
          $j=1:$ $(0\leqslant SOC< 0.4)$ & $1- SOC $ & 0 & 0 \\
          \hline
          $j=2:$ $(0.4 \leq SOC < 0.66 )$ & $SOC$ & \multirow{5}{*}{ $1- SOC $ }& \multirow{5}{*}{ $SOC $ }\\
          \cline{1-2}
          $j=3:$ $(0.66 \leq SOC < 0.73 )$ & $1- SOC $ &  & \\
          \cline{1-2}
          $j=4:$ $(0.73 \leq SOC < 0.75 )$ & $SOC$ &  & \\
          \cline{1-2}
         $j=5:$ $(0.75 \leq SOC < 0.86 )$ & $1- SOC $ &  & \\
          \cline{1-2}
          $j=6:$ $(0.86 \leq SOC\leqslant 1)$ & $SOC$ &  & \\
          \hline
      \end{tabular}
      \caption{SOC-dependent functions required for empirical correction of voltage prediction}
      \label{tab:ls}
  \end{table}

 \renewcommand{\arraystretch}{1.2}
  \begin{table}[h]
      \centering
      \begin{tabular}{|c|c|c|c|}
      \hline
          Region Switches $j-1 \to j$ & $m^{j|j-1}_{1}$ & $m^{j|j-1}_2$ & $m^{j|j-1}_3$ \\
          \hline
         $1\to 2$  & $SOC $ & $SOC $  & $SOC $  \\
          \hline
          $2 \to 3 $  & $1$ &  $1- SOC $ &  $SOC $ \\
          \hline
          $3 \to 4$  & $SOC -1$ &  $1- SOC $ &  $SOC $ \\
          \hline
          $4\to5$ & 1 & 0 & 0\\
          \hline
          $5\to 6$ & $1- SOC$ &  $SOC $ &  $SOC $ \\
          
          \hline
      \end{tabular}
      \caption{SOC-dependent functions required for empirical correction of error in voltage prediction}
      \label{tab:ms}
  \end{table}

\begin{figure}[h!]
    \centering
    \includegraphics[width=0.7\linewidth]{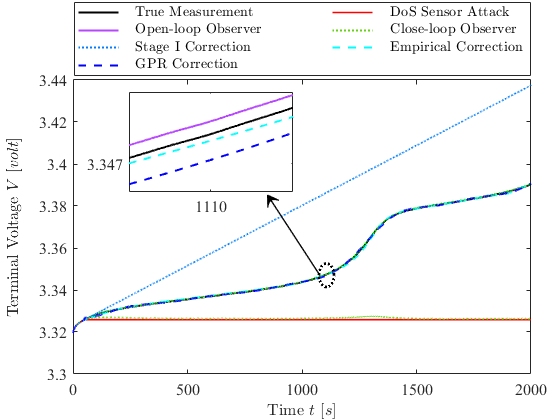}
    \caption{ \textbf{DoS sensor attack: }The plot shows the true and corrupt terminal voltage for the battery, the voltage estimation from open-loop and closed model-based observers, self-learning Koopman-based estimation with stage I correction only, with stage II empirical correction, and stage II GPR correction. }
    \label{fig:dos}
\end{figure} 

\noindent
\textbf{Denial-of-Service (DoS) sensor attack: } In this scenario, we consider that the adversary injects a DoS sensor attack at $50$s. Consequently, BMS ceases to receive the updated voltage measurement, rather continues to work with last terminal voltage measurement received before the  DoS attack.  This scenario is captured in Fig~\ref{fig:dos} that shows the  true and corrupted terminal respectively with black and red line.   Such attack can cause overcharging of the battery. The SOC at the  attack initiation is 0.36.  Under this DoS attack, both the empirical  and  GPR correction generates very accurate terminal voltage estimation which is particularly shown in the zoomed insect of Fig~\ref{fig:dos}. However, Fig~\ref{fig:dos} shows that the  Stage I corrections are unable to incorporate higher-order complexities of battery dynamics and continue to use the predictive slope from the last correct learning window as its basis for prediction. On the other hand, open-loop observer works well since the observer model is assumed to match the battery dynamics perfectly.   However, in real-world application this is unrealistic and we show the impact of model imperfection on open-loop observer estimates in our next case study. Closed-loop observers estimations based on attacked measurement expectedly follow the corrupt measurement and rendered are completely useless. This result underlines the need for our proposed strategy of secure estimation.

\begin{figure}[h!]
    \centering
    \includegraphics[width=0.7\linewidth]{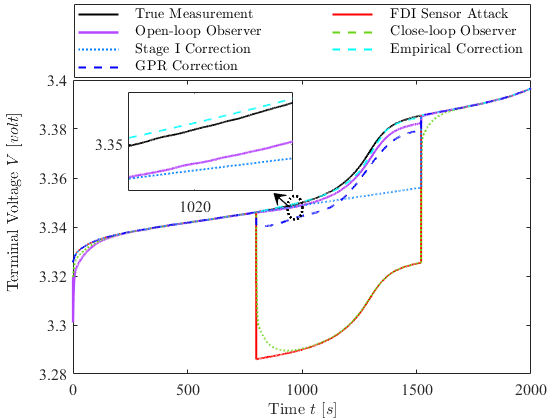}
    \caption{ \textbf{FDI sensor attack on aging battery:} The plot shows the true and corrupt terminal voltage for the battery, the voltage estimation from open-loop and closed model-based observers, self-learning Koopman-based estimation with stage I correction only, with stage II empirical correction, and stage II GPR correction. }
    \label{fig:fdi}
\end{figure} 

\noindent
\textbf{False-Data-Injection (FDI) sensor attack on aging battery:} We consider an aging battery for this case study. In particular, we consider increased increased internal resistance in ECM to capture aging effects following \cite{vilsen2019predicting}.   We further consider that an adversary launches a bias FDI of negative 0.06 volt on voltage measurement during the charging of this battery. The  attack starts at $80$s and continues for 12 minutes. Fig~\ref{fig:fdi} captures this scenario. The SOC at the beginning of this attack is 0.57. Under this attack, the empirical correction provides good voltage estimation as it learns the new model online (shown in the zoomed insect of Fig~\ref{fig:fdi}) . Conversely, GPR correction fails to generate accurate estimation as it needs retraining for aging batteries. As expected, open-loop observers also estimate the voltage poorly with the change in the battery model. However, an adaptive observer will remain infeasible in this scenario due to the compromised voltage measurement that is usually required to estimate varying system parameters. The closed-loop model follows the corrupt voltage measurement and stage I corrections follow the prediction slope from last learning window, as before.

\section{Conclusion} \label{conclu}
In this work, we have adopted a two-stage error correction integrated self-learning Koopman approach for secure terminal voltage estimation for Lithium-ion batteries subjected to malicious sensor attack during charging. In stage I error correction, we estimated the Koopman voltage prediction error to compensate for the error accumulation from the Koopman approximation. In stage II, we proposed two methods for the higher-order system dynamics corrections that remain unavailable to the self-learning feedback of the Koopman operator.  Our first method uses an empirical correction that exploits the OCV-SOC mapping to track the higher-order dynamics in the terminal voltage. Our proposed secure estimation scheme using empirical correction is interpretable, requires limited data and model knowledge. 
For our second error compensation strategy, we propose a GPR-based data-driven error correction that requires nominal terminal voltage measurements for pre-training. Our simulation case studies illustrate that the proposed secure estimator generates highly accurate voltage estimation under DoS and FDI sensor attacks. In addition, the empirical correction method reliably generates secure estimation for both new and aging battery without any re-training. Such secure estimation will enable the BMS to ensure optimal and safe charging of Lithium-ion battery while mitigating impact of sensor attack.


\bibstyle{arxiv}
\bibliography{ref.bib}

\end{document}